\documentstyle[twocolumn,11pt]{article}

\newcommand{\be}{\begin{equation}}
\newcommand{\ee}{\end{equation}}
\newcommand{\bea}{\begin{eqnarray}}

\newcommand{\eea}{\end{eqnarray}}

\def\gappeq{\mathrel{\rlap {\raise.5ex\hbox{$>$}}
{\lower.5ex\hbox{$\sim$}}}} 
\def\lappeq{\mathrel{\rlap{\raise.5ex\hbox{$<$}}
{\lower.5ex\hbox{$\sim$}}}}
 
\begin{document}

\date{}                                 %%%% so that date does not print. 

\title{{\bf Cell Microtubules as Cavities: Quantum Coherence and 
Energy Transfer? }}           
%%%% Replace with your title.

%%%%  Just replace the author names. 

\author{
N.E. Mavromatos \\
Department of Physics, Theoretical Physics Group, \\
University of London, King's College \\
Strand, London WC2R 2LS, United Kingdom. }

\maketitle                        %%%% To set Title and Author names.
\thispagestyle{empty}

%%%% Replace with your Abstract.

\noindent
{\bf Abstract
      {\small\em A model is presented for 
dissipationless energy transfer in cell 
microtubules due to quantum coherent states. The model is based on 
conjectured (hydrated) ferroelectric properties of microtubular
arrangements. Ferroelectricity is essential in providing the necessary 
isolation against thermal losses in thin interior regions, full of 
ordered water, near the tubulin dimer walls of the microtubule.
These 
play the role of cavity regions, which are similar to 
electromagnetic cavities of quantum optics.  
As a result, 
the formation of (macroscopic) quantum coherent states of electric
dipoles on the tubulin dimers may occur. 
Some experiments, inspired by quantum optics, are suggested
for the falsification of this scenario. }}

\vspace{0.5cm}


%%%%%%%%%%%%%%%%%%%%%%%%%%%%%%%%%%%%%%%%%%%%%%%%%%%%%%%%%%%%

\section{Introduction}

The r\^ole of Quantum Mechanics in Biological Matter 
is naively expected to be strongly suppressed, mainly due to 
the `macroscopic nature' of most biological entities
as well as the fact that biological matter 
lives in room temperature environments. 
These in general result in 
a very fast {\it collapse} of the pertinent 
wave-functions to a 
classical state. 
However, under certain circumstances it is possible
to obtain the necessary isolation against thermal losses and other
environmental entanglement, so that 
{\it macroscopic} quantum-mechanical 
coherence, 
extending over scales
that are considerably larger than the atomic scale, 
may be achieved and maintained
for some time, so that 
certain physical processes can occur. 
This is precisely the point of this presentation, namely we shall 
make an attempt to identify biological entities that can sustain 
quantum coherent states responsible for 
energy transfer without dissipation across the cell.

In particular, we shall argue that, under certain circumstances that we 
shall identify below, 
it is possible
for 
Cell MicroTubules (MT)~\cite{dustin}
to operate 
as quantum-mechanical isolated cavities,
exhibiting properties analogous to those of electromagnetic
cavities used in Quantum Optics~\cite{haroche}. 
The results presented here have been obtained in collaboration 
with D.V. Nanopoulos. In this short communication 
we shall only outline the 
main features. For details we refer the interested reader to ref. \cite{mn}.

Energy transfer across the cells,  
without dissipation,  
had been first conjectured to occur in 
biological matter 
by 
Fr\"ohlich~\cite{Frohlich}. 
The phenomenon conjectured by Fr\"ohlich was based 
on his one-dimensional superconductivity model:
in one dimensional electron systems with holes, 
the formation 
of solitonic structures due to electron-hole pairing
results 
in the transfer of electric 
current without dissipation.
In a similar manner, Fr\"ohlich conjectured 
that energy in biological matter could be transfered 
without dissipation, if appropriate solitonic 
structures are formed inside the cells.
This idea has lead theorists to construct 
various models for the energy transfer 
across the cell, based on the formation of kink 
classical solutions~\cite{lal}. 

In the early works no specific microscopic models had been 
considered~\cite{lal}.
However, after the identification of 
the MT as one of the most important structures of the cell,
both functionally and structurally, 
a model for their dynamics 
has been presented in ref. \cite{mtmodel}, 
in which the formation of solitonic structures,  
and their r\^ole in energy transfer across the MT, is discussed
in terms of classical physics.   
In ref. \cite{mn} we have considered the {\it quantum aspects}
of this one-dimensional model, and argued on the consistent 
quantization of the soliton solutions, as well as the fact that 
such semiclassical solutions may emerge as a result of `decoherence'
due to environmental entanglement, according to the 
ideas in ~\cite{zurek}.  

The basic assumption of the model used in ref. \cite{mn}
was that the fundamental structures
in the MT (more specifically of the brain MT) 
are Ising spin chains (one-space-dimensional structures).
The interaction of each chain with the neighbouring
chains and the surrounding water environment 
had been mimicked by suitable potential terms in the 
one-dimensional Hamiltonian.
The model describing the dynamics of such one-dimensional 
sub-structures 
was the ferroelectric distortive spin chain
model of ref. \cite{mtmodel}.

Ferroelecricity is an essential ingredient for the 
quantum-mechanical mechanism for energy transfer we discuss
here. We have conjectured~\cite{zioutas} that the ferroelectricity
which
might 
occur in MT arrangements, will be that of {\it hydrated} ferroelectrics,
i.e. the ordering of the electric dipoles will be due to 
the interaction of the dimers 
with the ordered-water molecules in the interior of 
the microtubular cavities. 
The importance of ferroelectricity lies on the fact that 
it induces a dynamical dielectric `constant'
$\varepsilon (\omega)$  which is 
dependent on the frequency $\omega$ of the excitations in the medium.
Below a certain frequency,  such materials 
are characterized by almost vanishing dynamical dielectric
`constants', which in turn implies that electrostatic 
interactions inversely proportional to $\varepsilon$ 
might be enhanced, and thus become dominant against thermal losses.
In the case of microtubules, the pertinent interactions 
are of electric dipole type, scaling 
with the distance $r$ as 
$1/(\varepsilon r^3)$. For ordinary water media, the dielectric constant 
is of order $80$. If ferroelectricity occurs, however, this is diminished 
significantly. As a result, the electric dipole-dipole interactions
may overcome the thermal losses $k_BT$ at room temperatures
inside an interior cylindrical region of MT bounded by the dimer walls 
of thickness of order of a few Angstroms~\cite{mn}. 

Once such an isolation is provided, it is possible for these thin 
interior regions to act as cavities in a way similar to 
that of quantum optical electromagnetic cavities~\footnote{Note that 
the role of MT as waveguides has been proposed by S. Hameroff
already some time ago~\cite{guides}. However, in our scenario
we look for isolated regions inside the MT which 
play the role of cavities.}. 
The latter are characterized by coherent modes 
of the electromagnetic radiation. In a similar spirit, 
one expects that such coherent cavity modes will occur 
in the above-mentioned thin interior regions of MT 
bounded by the protein dimer walls.
Indeed, as we discussed in \cite{mn}, these modes 
are provided by the interaction of the electric dipole moments
of the ordered-water molecules in the interior of MT with the 
quantised electromagnetic radiation~\cite{delgiud,prep}. Such 
coherent modes are termed {\it dipole quanta}. 

It is the interaction of such `cavity' modes with the 
electric dipole excitations of the dimers that leads to 
the formation of coherent (dipole) states on the tubulin dimer
walls of MT. A review of how this can happen, 
and by what means one can 
test such a situation experimentally, will be the main topic 
of this communication.

\section{Microtubules and Coherent States} 

Microtubules 
are hollow cylinders 
comprised of an exterior surface
of cross-section
diameter
$25~nm$ ($1~nm=10^{-9}$ meters) 
with 13 arrays
(protofilaments)
of protein
dimers
called tubulines~\cite{dustin}.
The interior of the cylinder,
of cross-section
diameter $14~nm$,
contains {\it ordered water} molecules,
which implies the existence
of an electric dipole moment and an electric field.
We stress that the {\it ordered water} 
may be different from ordinary water.
The arrangement of the dimers is such that, if one ignores
their size,
they resemble
triangular lattices on the MT surface. Each dimer
consists of two hydrophobic protein pockets, and
has an unpaired electron.
There are two possible positions
of the electron, called $\alpha$ and $\beta$ {\it conformations}. 
When the electron is
in the $\beta$-conformation there is a $29^o$ distortion
of the electric dipole moment as compared to the $\alpha $ conformation.

In standard models for the simulation of the MT dynamics~\cite{mtmodel},
the `physical' degree of freedom -
relevant for the description of the energy transfer -
is the projection of the electric dipole moment on the
longitudinal symmetry axis (x-axis) of the MT cylinder.
The $29^o$ distortion of the $\beta$-conformation
leads to a displacement $u_n$ along the $x$-axis,
which is thus the relevant physical degree of freedom.
This way, the effective system is one-dimensional (spatial),
and one has the possibility of a quantum integrable
system~\cite{mn}. 

Information processing
occurs via interactions among the MT protofilament chains.
The system may be considered as similar to a model of
interacting Ising chains on a triangular lattice, the latter being
defined on the plane stemming from filleting open and flattening
the cylindrical surface of MT.
Classically, the various dimers can occur in either $\alpha$
or $\beta$ conformations. Each dimer is influenced by the neighbouring
dimers resulting in the possibility of a transition. This is
the basis for classical information processing, which constitutes
the picture of a (classical) cellular automatum.

The {\it quantum nature} of a MT network results
from the {\it assumption} that each dimer finds itself in a
{\it superposition} of $\alpha$ and $\beta$ conformations.
Viewed as a {\it two-state} quantum mechanical 
system, the MT tubulin dimers couple to conformational changes with
$10^{-9}-10^{-11} {\rm sec}$ transitions, corresponding to an
angular frequency $     \omega \sim{\cal O}( 10^{10})
-{\cal O}(10^{12})~{\rm Hz}$. In the present work we 
assume the upper bound of this frequency range 
to represent (in order of magnitude) the characteristic frequency 
of the dimers, viewed as a two-state quantum-mechanical system: 
\be
     \omega _0 \sim {\cal O}(10^{12})~{\rm Hz} 
\label{frequency2}
\ee
As we shall see below, such frequency range is not unusual 
for biological matter. 

Let $u_n$ be the displacement field of the $n$-th dimer in a MT
chain.
The continuous approximation proves sufficient for the study of
phenomena associated with energy transfer in biological cells,
and this implies that one can make the replacement
\be
  u_n \rightarrow u(x,t)
\label{three}
\ee
with $x$ a spatial coordinate along the longitudinal
symmetry axis of the MT. There is a time variable $t$
due to
fluctuations of the displacements $u(x)$ as a result of the
dipole oscillations in the dimers.

The effects of the neighbouring
dimers (including neighbouring chains)
can be phenomenologically accounted for by an effective
potential $V(u)$. In the model of ref. \cite{mtmodel} 
a double-well potential was used, leading to a 
classical kink solution for the $u(x,t)$ 
field. More complicated interactions are allowed
in the picture of ref. \cite{mn}, where 
more generic polynomial potentials have been considered.

The effects of the surrounding water molecules can be
summarised by a viscous force term that damps out the
dimer oscillations,
\be
 F=-\gamma \partial _t u
\label{six}
\ee
with $\gamma$ determined phenomenologically at this stage.
This friction should be viewed as an environmental effect, which
however does not lead to energy dissipation, as a result of the
non-trivial
solitonic structure of the
ground-state
and the non-zero constant
force due to the electric field.
This is a well known result, directly relevant to
energy transfer in biological systems \cite{lal}.

In mathematical terms the effective equation of motion 
for the relevant field degree of freedom $u(x,t)$ reads:
\be
u''(\xi) + \rho u'(\xi) = P(u) 
\label{generalsol}
\ee
where $\xi=x-vt$, $v$ is the velocity of the soliton, 
$\rho \propto \gamma$~\cite{mtmodel}, 
and 
$P(u)$ is a polynomial in $u$, of a certain degree, stemming 
from the variations of the potential $V(u)$ describing interactions
among the MT chains~\cite{mn}.
In the mathematical literature~\cite{otinowski} 
there has been a classification of solutions of equations
of this form. For certain forms of the potential~\cite{mn} the 
solutions
include {\it  kink solitons} that may be 
responsible for dissipation-free energy transfer in biological
cells~\cite{lal}: 
\begin{equation} 
u(x,t) \sim c_1 \left(tanh[c_2(x-v t)] + c_3 \right)
\label{kink}
\end{equation}
where $c_1,c_2, c_3$ are constants depending on the 
parameters of the dimer lattice model.
For the form of the potential assumed in the model of \cite{mtmodel} 
there are solitons of the form 
$ u(x,t)= c_1' + \frac{c_2' - c_1'}{1 + e^{c_3'(c_2'-c_1')(x - vt)}}$,
where again $c_i', i=1,\dots 3$ are appropriate constants. 

A semiclassical quantization of such solitonic states
has been considered in \cite{mn}. 
The result of such a quantization yields a
modified soliton equation for the (quantum corrected) field
$u_{q}(x,t)$ \cite{tdva}
\be
    \partial ^2 _t u_q(x.t) - \partial _x ^2 u_q(x,t)
+ {\cal M}^{(1)} [u_q(x,t)] = 0
\label{22c}
\ee
with the notation \\
$M^{(n)} = e^{\frac{1}{2}(G(x,x,t)-G_0(x,x))\frac{\partial ^2}
{\partial z^2}} U^{(n)}(z) |_{z=u_q(x,t)}$,
and $U^{(n)} \equiv d^n U/d z^n$. 
The quantity $U $ denotes the potential of the original soliton
Hamiltonian,
and $G (x,y,t)$ is a bilocal field that describes quantum corrections
due to the modified
boson field around the soliton.
The quantities $M^{(n)}$ carry information about the
quantum corrections.
For the kink soliton (\ref{kink}) the quantum
corrections (\ref{22c})
have been calculated explicitly in ref. \cite{tdva},
thereby providing us with a  concrete
example of a large-scale quantum coherent state.

A typical propagation velocity of the kink solitons (e.g. in the model of 
ref. \cite{mtmodel})
is $v \sim 2~{\rm m/sec}$, although, models with $v \sim 20~m/sec$ 
have also been considered~\cite{satar}. 
This implies that, 
for moderately long microtubules 
of length $L \sim 10^{-6}$ m, such kinks transport energy 
without dissipation 
in 
\be
      t_F \sim 5 \times 10^{-7}~{\rm sec}
\label{FS}
\ee
As we shall see in the next section, such time scales
are comparable to, or smaller in magnitude than, the 
decoherence time scale of the above-described  coherent (solitonic) 
states $u_q (x,t)$.
This implies the possibility that fundamental quantum 
mechanical phenomena may then be responsible for frictionless
energy (and signal) transfer across microtubular arrangements in the cell.

\section{Microtubules as Cavities}

In ref. \cite{mn} we have presented a microscopic 
analysis of the physics underlying the interaction
of the water molecules with the dimers of the MT,
which is responsible for providing the friction term (\ref{six})
in the effective (continuum) description.
Below we review briefly the scenario.
As a result of 
the ordered structure of the water environment in the interia of MT,
there appear {\it collective} coherent modes, 
the so-called dipole quanta~\cite{delgiud}. These arise from the 
interaction of the electric dipole
moment  of the water molecule with the quantised radiation
of the electromagnetic field~\cite{prep}, 
which may be self-generated in the case 
of MT arrangements~\cite{satar,mn}. 
Such coherent modes play the role of `cavity modes'
in the quantum optics terminology. 
These
in turn interact with the dimer structures, mainly through 
the unpaired electrons of the dimers,
leading to the formation of a quantum coherent 
solitonic state that may extend even over the entire MT network. 
As mentioned above, such states may be identified~\cite{mn} 
with semi-classical solutions of the 
friction equations (\ref{generalsol}). 
These coherent, 
almost classical, states
should be viewed as the result of {\it decoherence} of the 
dimer system due to its interaction/coupling with the 
water environment~\cite{zurek}.

Such a dimer/water coupling can lead 
to a situation analogous to 
that of
atoms interacting with coherent modes 
of the electromagnetic radiation in 
{\it quantum optics Cavities}, namely to the
so-called {\it Vacuum-Field Rabi Splitting} (VFRS) effect~\cite{rabi}. 
VFRS appears in both the emission and absorption~\cite{agar} 
spectra of atoms
in interaction with a coherent mode of electromagnetic radiation in a 
cavity. For our purposes below we shall 
review the phenomenon by 
restricting ourselves
for definiteness to the absorption spectra case.

Consider a collection of $N$ atoms of frequency $\omega_0$ inside an
electromagnetic cavity, to which it is injected a pulse 
of an external field, of frequency $\Omega$. 
Then, 
there is a doublet structure (splitting) of the absorption spectrum
of the atom-cavity system 
with peaks 
at:
\be
\Omega = \omega _0 - \Delta/2 \pm \frac{1}{2}( \Delta ^2 + 
4 N \lambda ^2 )^{1/2}
\label{rabiabs}
\ee
where $\Delta = \omega_c - \omega_0 $ is the detuning of the cavity mode, 
of frequency $\omega_c$,  
compared to the atom
frequency. 
For resonant cavities the splitting occurs with equal weights
\be
  \Omega = \omega_0 \pm \lambda \sqrt{N} 
\label{rabisplitting}
\ee
Notice here the {\it enhancement} of the effect 
for multi-atom systems $N >> 1$. 
The quantity  $2\lambda \sqrt{N}$ is called the `Rabi frequency'~\cite{rabi}. 
{}From the emission-spectrum theoretical analysis 
an estimate of $\lambda$ may 
be inferred which involves 
the matrix element, ${\underline d}$, of atomic electric dipole 
between the energy states
of the two-level atom~\cite{rabi}: 
\be 
   \lambda = \frac{E_{c}{\underline d}.{\underline \epsilon}}{\hbar}
\label{dipolerabi}
\ee
where ${\underline \epsilon}$ is the cavity (radiation)  mode 
polarisation, and 
\be
E_{c} \sim  \left(\frac{2\pi \hbar \omega_c}{\varepsilon  V}\right)^{1/2} 
\label{amplitude}
\ee
is the r.m.s. vacuum (electric) field amplitude at the centre 
of a cavity of volume $V$, and of frequency $\omega_c$,
with $\varepsilon $ the 
dielectric constant of the medium inside the volume $V$. 
In atomic physics 
the VFRS effect has been confirmed
by 
experiments 
involving beams of Rydberg atoms resonantly 
coupled to superconducting cavities~\cite{rabiexp}. 

Under the analogy between the thin cavity regions near the dimer
walls of MT with the above-mentioned electromagnetic cavities,
the role of atoms in this case is played by the unpaired
two-state electrons of the tubulin dimers~\cite{mn}
oscillating with a  frequency (\ref{frequency2}). 
To estimate the Rabi coupling between cavity modes and 
such dimer oscillations, one should use (\ref{dipolerabi}) 
for the MT case. 
The electric dipole moment $d$ in that case may be estimated
as follows: each dimer has a mobile 
charge~\cite{dustin}: $q=18 \times 2e$, $e$ the electron charge,
and we 
use the fact that a typical distance for the estimate 
of the electric dipole
moment for the `atomic' transition between the $\alpha,\beta$
conformations is 
of order of the 
distance between the two hydrophobic dimer pockets, i.e. 
${\cal O}(4~{\rm nm})$. 
This yields 
\be
d_{dimer} \sim  3 \times 10^{-18}~ {\rm Cb} \times {\rm Angstrom} 
\label{dipoledimer}
\ee
We also took account of the fact that, as a result of the 
water environment, the electric charge of the dimers appears
to be screened by the relative 
dielectric constant of the water, which 
in units of that of the vacuum is $\varepsilon/\varepsilon_0 \sim 80$. 
We note, however, that the biological environment of the unpaired 
electric charges in the dimer may lead to 
further suppression of $d_{dimer}$  (\ref{dipoledimer}).

We have used some 
simplified models for the ordered-water molecules, 
which yield~\cite{mn} 
a frequency of the coherent dipole quanta (`cavity' modes)
of order: 
\be
     \omega_c \sim 6 \times 10^{12} s^{-1} 
\label{frequency}
\ee
Notably this is of the same order 
as
the characteristic frequency of the dimers
(\ref{frequency2}), implying that  
the dominant cavity mode and the dimer system are almost in resonance
in the scenario of \cite{mn}.
Note that this is a feature shared by  
the Atomic Physics systems in Cavities, and thus 
we may apply the pertinent formalism to our system.
Assuming a relative dielectric constant of water 
w.r.t to that of vacuum $\epsilon_0$, $\varepsilon/\varepsilon_0 \sim 80$, 
one obtains from (\ref{amplitude}) for the case of MT cavities:
\be
    E_{c} \sim 10^{4}~{\rm V/m}
\label{eowmt}
\ee 
Notably, electric fields of such a magnitude 
can be provided by the electromagnetic interactions 
of the MT dimer chains, the latter 
viewed as giant electric dipoles~\cite{mtmodel}. 
This may be seen to suggest that the 
coherent modes $\omega_c$, which in our scenario 
interact with the unpaired electric charges of the dimers 
and produce the kink solitons along the chains, 
owe their existence to  the 
(quantised) electromagnetic interactions 
of the dimers themselves.

The 
Rabi coupling for the MT case then is estimated 
from (\ref{dipolerabi}) to be of order: 
\bea 
&~&    {\rm Rabi~coupling~for~MT} \equiv \lambda _{MT} 
= \nonumber \\
&~& \sqrt{{\cal N}} \lambda_0 \sim 3 \times 10^{11} s^{-1} 
\label{rabiMT}
\eea
which is, on average, an order of magnitude 
smaller than the characteristic frequency 
of the dimers (\ref{frequency2}). 

In the above analysis, it was assumed that the system of MT dimers 
interacts with a {\it single} dipole-quantum coherent 
mode of the ordered 
water and hence interactions among the dimer 
spins were ignored. More complicated situations, involving 
interactions either 
among the dimers or of the dimers with more than 
one radiation quanta,
which undoubtedly occur
in nature, may affect the above estimate.  

The presence of such a coupling between water molecules and dimers
leads to quantum coherent solitonic states of electric dipole
quanta on the tubulin dimer walls. 
To estimate the decoherence time 
we remark that the main source 
of dissipation (environmental entanglement) comes from the 
imperfect walls of the cavities, which lose coherent modes
and energy. 
The time scale, $T_r$, over which a cavity MT dissipates its energy,  
can be identified in our model with the average life-time $t_L$  
of a coherent-dipole quantum state,
which has been found to be~\cite{mn}: 
$T_r \sim t_L \sim  10^{-4}~{\rm sec}$.
This leads to a naive estimate 
of the quality factor for the MT cavities,
$Q_{MT} \sim \omega_c T_r \sim {\cal O}(10^8)$. 
We note, for comparison, that high-quality 
cavities encountered in Rydberg atom
experiments 
dissipate energy in time scales of ${\cal O}(10^{-3})-{\cal O}(10^{-4})$
sec, and have $Q$'s which are comparable to $Q_{MT}$ above. 
The analysis of 
~\cite{mn} then yields the following estimate 
for the collapse time of the kink coherent state of the MT dimers 
due to dissipation:
\be
t_{collapse} \sim {\cal O}(10^{-7})-{\cal O}(10^{-6})~{\rm sec}
\label{tdecohsoliton}
\ee
This is larger than the time scale (\ref{FS})
required for energy transport 
across the MT by an average kink soliton in the models of  
\cite{mtmodel,satar}. The result (\ref{tdecohsoliton}), then, 
implies that 
Quantum Physics may not be
irrelevant as far as 
dissipationless energy transfer across the MT is concerned. 

In view of this specific model, 
we are therefore in disagreement at this stage 
with the 
conclusions of \cite{tegmark}, that only classical physics
is responsible for energy and signal transfer in brain MT.
Such conclusions did not take proper account 
of the possible isolation against environmental 
entanglement, which may occur inside certain regions of MT.

\section{Conclusions and Outlook}

In \cite{mn} we have put forward a conjecture
concerning the representation of certain regions 
inside the MT arrangements 
in the cell as {\it isolated} high-Q(uality) {\it cavities}. 
We presented a scenario according to which 
the presence of the ordered water in the interior of the MT cylindrical 
arrangements results in the appearance of electric dipole quantum 
coherent modes, which couple to the unpaired electrons of the 
MT dimers via Rabi vacuum field couplings.
The situation is analogous to  
the physics of Rydberg atoms in electromagnetic cavities~\cite{rabi}. 
In quantum optics, such couplings may be considered as experimental 
proof of the quantised nature of the electromagnetic radiation.
In our case, therefore, if present, such couplings could 
indicate the existence of the coherent quantum modes
of electric dipole quanta in the ordered water environment of MT,
conjectured in ref. \cite{delgiud,prep}, and used here.

To verify experimentally such a situation, 
one should first check on the ferroelectric properties
of the MT arrangements, which the above analysis is 
crucially based on, 
and measure the corresponding dipole moments 
of the tubulin dimers. 
A suggestion along these lines has been put forward in 
ref. \cite{zioutas}. 

In addition, 
one should verify 
the aforementioned Vacuum field Rabi coupling (VFRS),
$\lambda_{MT}$,  
between the MT dimers and the ordered water quantum coherent modes. 
This coupling, if present,  
could be tested 
experimentally  
by the same methods used to measure VFRS in atomic physics~\cite{rabiexp}, 
i.e. by using the MT themselves as {\it cavity environments}, 
in the way described above, 
and considering tunable probes to excite the coupled dimer-water
system. Such probes could be pulses of (monochromatic) light, for example,
passing through the hollow cylinders of the MT. 
This would be the analogue of an external field in the atomic experiments
mentioned above. The field would then resonate, not at the bare frequencies 
of the coherent dipole quanta or dimers, but at the {\it Rabi splitted} ones,
leading to a double peak in the absorption 
spectra of the dimers~\cite{rabiexp}. By using MT of different sizes
one could thus check on 
the characteristic $\sqrt{N}$-enhancement of the (resonant) Rabi coupling 
(\ref{rabisplitting}) for MT systems with $N$ dimers. 

Clearly much more work needs to be carried out before even tentative
conclusions are reached, concerning the nature of the MT 
arrangements inside the cell. Moreover, one should always bear in mind that
the extrapolation of results obtained 
{\it in vitro} to {\it in vivo} situations may not always be feasible. 
However, we believe that the above theoretical ideas and 
experimental suggestions 
constitute a useful addition to the programme of understanding 
the dynamics of Microtubules, and the associated
processes of energy and signal (stimuli) transfer across the cells.

\end{document}